\documentclass[showpacs,preprintnumbers,amsmath,amssymb,prb]{revtex4}

\usepackage{graphicx}% Include Fig.files
\usepackage{dcolumn}% Align table columns on decimal point
\usepackage{bm}% bold math
\usepackage{amsfonts}
\usepackage{amssymb}

\newcommand{\no}[1]{}

\def\drawline#1#2{\raise 2.5pt\vbox{\hrule width #1pt height #2pt}}

\def\trian{\raise 1.25pt\hbox{$\scriptscriptstyle\triangle$}\nobreak\ }

\def\square{${\vcenter{\hrule height .4pt
        \hbox{\vrule width .4pt height 3pt \kern 3pt
        \vrule width .4pt}
        \hrule height .4pt}}$\nobreak\ }

\def\plus{\raise 1.25pt \hbox{$\scriptscriptstyle +$}\nobreak\ }

\begin{document}

\title{Moist turbulent Rayleigh-B\'{e}nard convection with Neumann and Dirichlet boundary conditions}

\author{Thomas Weidauer}
\affiliation{Institut f\"ur Thermo- und Fluiddynamik, Technische Universit\"at Ilmenau, Postfach 100565, D-98684 Ilmenau, Germany}
\author{J\"org Schumacher}
\affiliation{Institut f\"ur Thermo- und Fluiddynamik, Technische Universit\"at Ilmenau, Postfach 100565, D-98684 Ilmenau, Germany}

\date{\today}

\begin{abstract}
Turbulent Rayleigh-B\'enard convection with phase changes in an extended layer between two parallel impermeable planes is studied by means of three-dimensional direct numerical simulations for Rayleigh numbers between $10^4$ and $1.5\times 10^7$ and for Prandtl number $Pr=0.7$. Two different sets of boundary conditions of temperature and total water content are compared: imposed constant amplitudes which translate into Dirichlet boundary conditions for the scalar field fluctuations about the quiescent diffusive equilibrium and constant imposed flux boundary conditions that result in Neumann boundary conditions. Moist turbulent convection is in the conditionally unstable regime throughout this study for which unsaturated air parcels are stably and saturated air parcels unstably stratified. A direct comparison of both sets of boundary conditions with the same parameters requires to start the turbulence simulations out of differently saturated equilibrium states. Similar to dry Rayleigh-B\'{e}nard convection the differences in the turbulent velocity fluctuations, the cloud cover and the convective buoyancy flux decrease across the layer with increasing Rayleigh number. At the highest Rayleigh numbers the system is found in a two-layer regime, a dry cloudless and stably stratified layer with low turbulence level below a fully saturated and cloudy turbulent one which equals classical Rayleigh-B\'enard convection layer. Both are separated by a strong inversion that gets increasingly narrower for growing Rayleigh number.
\end{abstract}

\maketitle

\section{Introduction}
Turbulent convection is omnipresent in the atmosphere of the Earth\cite{Atkinson,Wyngaard} and other planets.\cite{Sanchez} In many cases the convective motion is accompanied by phase changes. When warm air parcels rise they cool down and can form a condensate or in other words -- a cloud. Moist convection combines thus turbulent motion of characteristic plumes as known from the classical dry convection with phase transitions between vapor and liquid water (or even ice as a third phase). The dynamics of clouds contains a wide range of spatial and temporal scales starting from cloud microphysical processes on millimeter scales within seconds up to the dynamics of whole clouds or even cloud clusters on scales larger than kilometers with lifetimes of an hour or more.\cite{Blyth} On all scales turbulence interacts with the cloud dynamics and the phase changes in multiple ways which are not yet fully understood, but have to be parametrized in global models of atmospheric motion.\cite{Siebesma,Stevens,Bodenschatz2010} Progress can be achieved in two ways. Either one increases the number of physical processes and thus the complexity of the parametrizations as necessary in large-eddy simulations (LES) or cloud-resolving models to meet the variety of cloud phenomena \cite{Siebesma,Siebesma2} or one tries to reduce the dynamics to a few essential building blocks and disentangles their complex interplay. The latter route is taken in the present work.

In this paper, we study therefore moist Rayleigh-B\'{e}nard convection (MRBC) with simplified thermodynamics of the phase changes. The model equations will make use of the Boussinesq approximation which means that variations of thermodynamic state variables about their reference profiles remain small. The model describes thus aspects of shallow moist convection as present in low cumulus \cite{Siebesma} and stratocumulus clouds.\cite{Stevens} Oceanic and continental surfaces at the bottom of the troposphere have different thermal properties and can act differently on the turbulent convection. This in turn can directly affect the cloud formation as pointed out e.g. by Bretherton.\cite{Bretherton2} Convection over land surfaces supplies usually very heterogenous surface conditions, such as different degrees of roughness or moisture fluxes due to changes in vegetation. Such effects were studied in LES by a reduction of complex surface conditions to a strip-like heterogeneity with a prescribed variation scale.\cite{Moeng2005} It was shown that the resulting buoyancy fluxes depend sensitively on the heterogeneity scale due to partly initiated secondary convective circulations close to the boundary. In the simplest case, boundary conditions (BC) for the temperature and moisture can be prescribed by constant values, constant fluxes or mixtures of both. In particular, constant amplitude boundary conditions can be considered as a first approximation of moist convection over an ocean surface with a constant sea surface temperature, constant flux boundary conditions as a first approximation for a shelf region
close to the coast line.

The present work will study the impact of such simple boundary conditions on the buoyancy flux and the cloud formation by means of three-dimensional direct numerical simulations (DNS). How do the differences in the turbulent transport for both BC depend on the Rayleigh number? The atmospheric context as a motivation in mind, we will carry out the DNS in flat cells with an aspect ratio of 16 and larger. Our MRBC model is a direct extension of the classical dry Rayleigh-B\'{e}nard convection incorporating the physics of phase changes in a simplified way, but keeping the main ingredient, the release of latent heat (for more details, see Sec. II and Ref. \onlinecite{Olivier1}). This simplification makes the present model accessible to parametric DNS studies without subgrid-scale parametrizations of turbulence. In a series of DNS studies with our MRBC model, which used so far the constant value BC only, we demonstrated that some aspects of the complex atmospheric cloud dynamics can be reproduced qualitatively well by the present model.\cite{Schumacher1,Weidauer1} This holds particularly for the so-called {\em conditionally unstable} regime of moist convection, in which dry unsaturated air is stably  and moist saturated air unstably stratified with respect to small vertical lifts of the air parcels.\cite{Weidauer2,Olivier2} Moist convection is known to form then localized cloud aggregates in which the moist air rises up and is surrounded by ambient unsaturated regions with a significantly reduced level of turbulent fluctuations.\cite{Bretherton1,Bretherton2,Olivier2} Similar to wall-bounded shear flows, conditionally unstable moist convection is triggered by a finite-amplitude perturbation of the (linearly stable) equilibrium.\cite{Weidauer2} This is a situation that can exist when cumulus clouds are formed.\cite{Siebesma}

Interestingly in classical dry Rayleigh-B\'enard convection the case of fixed temperatures at the plates is mostly investigated.\cite{Kadanoff,Ahlers,Lohse} Only in the last years fixed flux BC or combinations of fixed value and fixed flux BC came into the focus of interest in order to understand differences in simulations and experiments. Emphasis was given to the dependence of the global turbulent heat transport as quantified by the Nusselt number $Nu$ on the Rayleigh number $Ra$, one input parameter of convection.\cite{Verzicco,Johnston} Otero et al.\cite{Otero} and later Wittenberg\cite{Wittenberg} derived an analytical upper bound on the turbulent heat transport for the case of fixed flux and no-slip boundary conditions for the flow that has the same dependence on the Rayleigh number as the fixed temperature case, that is $Nu \le const \times Ra^{1/2}$. It is known that constant flux BC decrease the critical Rayleigh number $Ra_c$ for the onset of convection and shift instabilities to the largest scales.\cite{Hurle,Chapman} The critical wavenumber is thus zero in an infinitely extended layer. This also holds for stress-free or free-slip BC of the velocity field. It is different to the fixed-temperature case where the critical wavenumber is of the order of one. Thus for Rayleigh numbers right above the linear instability threshold a box of arbitrary length will contain a single convection roll only and we can expect that relics of this property are observable in the turbulent regime.

The impact of fixed flux BC on turbulent heat transport in Rayleigh-B\'{e}nard convection is still a matter of ongoing research. In Ref. \onlinecite{Verzicco}, DNS with fixed heat flux BC at the lower heating plate and fixed temperature at the upper cooling plate of a cylindrical cell were performed. It was found that for Rayleigh numbers larger than $10^8$ the Nusselt number $Nu$ is larger than for the case of both plates being held at constant temperature. Differences in the evolution of the rising thermal plumes at the lower plate were suggested as a reason for such an increase. Later it was shown that for higher resolutions the plume detachment does not depend on the particular choice of temperature boundary conditions.\cite{Stevens2} A further series of  DNS was done by Johnston and Doering.\cite{Johnston} The authors compared both sets of BC within two-dimensional simulations of Rayleigh-B\'enard convection with periodic side walls over a large range of Rayleigh numbers. Their work was focussed on the turbulent heat transport and its dependence on the aspect ratio of the domain. The main result was that for Rayleigh numbers increasing beyond $Ra\sim10^7$ the Nusselt number of the fixed flux case converges to that of fixed temperature case and eventually coincides for $10^8\le Ra\le 10^{10}$. The authors mention that two-dimensional convective turbulence could be however different to the three-dimensional case. DNS in three dimensions by Hunt et al. \cite{Hunt2003} compared the evolving plume structures and temperature fluctuations for both sets of boundary conditions. Fluctuations were found to be enhanced when the constant flux case is compared with the constant temperature case. This previous work on dry Rayleigh-B\'{e}nard convection together with the original motivation coming from applications in an atmospheric context sets the stage for our present study.

The outline of this paper is as follows. The assumptions and model equations of moist Rayleigh-B\'{e}nard convection, the parameter regime and the numerical implementation are explained in Sec. \ref{Abschnitt2}. It is discussed how runs with the different sets of BC have to be compared to each other. Section \ref{Abschnitt3} presents the results of the DNS starting with reference runs which we performed without phase transitions. It is followed by results of the runs that include phase transitions. Here, we report the dependence of mean profiles, vertical fluxes and velocity fluctuations on the Rayleigh number. Finally, we will give a short summary and outlook to future work.

\section{\label{Abschnitt2}Moist convection model}

\subsection{Equations of motion and boundary conditions}
In this chapter a short introduction to the moist Rayleigh-B\'{e}nard convection model is given. It is based on works of Kuo \cite{Kuo} and Bretherton \cite{Bretherton1,Bretherton2} who studied the effect of latent heat release and phase changes on the structure and organization of moist convection at small Rayleigh numbers. A detailed derivation of our model can be found in Ref. \onlinecite{Olivier1}. The model combines the classical dry Rayleigh-B\'{e}nard convection with condensation and evaporation of water. The main assumptions of the model are as follows. We consider conversions between water vapor and liquid water only and exclude the third phase of ice. The air parcels are assumed to be in local thermodynamic equilibrium, i.e., the vapor content $q_v$ and the liquid water content $q_l$ can be summed up to the total water content $q_T$. The present model incorporates the Boussinesq approximation which limits it to convection in shallow layers. It is also used that in the vicinity of the phase boundary the buoyancy $B$ of an air parcel at a height $z$ can be expressed as a {\em linear} combination of the two thermodynamic state variables, the entropy (or potential temperature) $S$ and the total water content $q_T$. This step preserves the discontinuity of partial derivatives of $B(S,q_T,z)$ at the phase boundary which are given by
%---------------------------------------------------------------------------
\begin{eqnarray}
\frac{\partial B} {\partial S} \Big|_{q_T,z}&=& \left\{ \begin{array}{l l} B_{S,u}  & \textrm{ if } q_T \le q_{sat}(S,z) \\
B_{S,s}  & \textrm{ if } q_T > q_{sat}(S,z) \end{array}  \right. \label{BS}  \\
\frac{\partial B} {\partial q_T}\Big|_{S,z} &=& \left\{ \begin{array}{l l} B_{q_T,u} & \textrm{ if } q_T \le q_{sat}(S,z)  \\
B_{q_T,s} & \textrm{ if } q_T > q_{sat}(S,z)\,, \end{array} \right. \label{BQ}
\end{eqnarray}
%----------------------------------------------------------------------------
and thus the release of latent heat. Note that the constants differ, i.e., $B_{S,u}\ne B_{S,s}$ and $B_{q_T,u}\ne B_{q_T,s}$. Since the thermodynamic quantities are linearized on both sides of the phase boundary they can be combined to two new prognostic state variables, a \emph{dry} buoyancy field $D$ and a \emph{moist} buoyancy field $M$,
%--------------------------------------------------------------------------
\begin{eqnarray}
{D}&=& B_{S,u} ({S} - {S}_{ref})+ B_{q_T,u}({q}_T - {q}_{T,ref})  \\
{M}&=& B_{S,s} ({S} - {S}_{ref})+ B_{q_T,s}({q}_T - {q}_{T,ref})\,.
\end{eqnarray}
%-------------------------------------------------------------------------
The dry buoyancy ${D}$ can be interpreted to be proportional to the so-called liquid water potential temperature and the moist buoyancy ${M}$ to be proportional to the equivalent potential temperature. We have thus converted the buoyancy ${B}({q}_T,{S},{z})$ into ${B}({D},{M},{z})$ which is determined for each space-time point by \cite{Olivier1}
%--------------------------------------------------------------------------
\begin{equation}
{B}({\bf {x}},{t})=\max [{M}({\bf {x}},{t}), {D}({\bf {x}},{t})-N_s^2 {z}]\,, \label{buoorig}
\end{equation}
%--------------------------------------------------------------------------
where $N_s$ is the Brunt-Vaisala frequency which is composed of the dry and moist adiabatic lapse rates. We use the common practice to decompose both buoyancy fields $D$ and $M$ into linear 
background profiles ${\bar{D}}({z})$ and ${\bar{M}}({z})$ which determine the quiescent diffusive equilibrium of the convection layer and fluctuating parts ${D}^\prime$ and ${M}^\prime$. The linear profiles are defined via the buoyancy amplitudes at the lower and upper planes which are denoted by ${D}_0$, ${D}_H$, ${M}_0$ and ${M}_H$. For the dry and moist buoyancies this gives
%--------------------------------------------------------------------------
\begin{eqnarray}
\label{EquiD} {{\bar{D}}}({z})&=& {D}_0 +\frac{{D}_H-{D}_0}{H} {z}\,,\\
\label{EquiM} {{\bar{M}}}({z})&=& {M}_0 +\frac{{M}_H-{M}_0}{H} {z}\, .
\end{eqnarray}
%--------------------------------------------------------------------------
To obtain dimensionless equations one has to define characteristic scales. These are the height of the layer $H$, the characteristic moist buoyancy difference between top and bottom planes ${M}_0-{M}_H$,  the free-fall velocity $U_f=\sqrt{({M}_0-{M}_H)H}$, the resulting time scale $T_f=H/U_f$ and the characteristic kinematic pressure $U_f^2$. The dimensionless Navier-Stokes equations in the Boussinesq approximation for shallow moist convection are then given by (in this chapter we indicate dimensionless quantities with a tilde)
%--------------------------------------------------------------------------
\begin{eqnarray}
\frac{\partial \tilde{\bf u}}{\partial \tilde{t}} +( \tilde{\bf u} \cdot \tilde{\nabla} ) \tilde{\bf u} &=& -\tilde{\nabla} \tilde{p} + \sqrt{\frac{Pr}{Ra_M}} 
\tilde{\nabla}^2 \tilde{\bf u} + \tilde{B} {\bf e}_z \label{GG7} \\
\tilde{\nabla} \cdot \tilde{\bf u} &=& 0  \label{GG1} \\
\frac{\partial \tilde{D}^\prime}{\partial \tilde{t}} + ( {\bf \tilde{u}} \cdot \tilde{\nabla} ) \tilde{D}^\prime &=& \frac{1}{\sqrt{Pr Ra_M}} \tilde{\nabla}^2 
\tilde{D}^\prime +\frac{Ra_D}{Ra_M} \tilde{u}_z \label{GG3} \\
\frac{\partial \tilde{M}^\prime}{\partial \tilde{t}} + (\tilde{\bf u} \cdot \tilde{\nabla} ) \tilde{M}^\prime &=& \frac{1}{\sqrt{Pr Ra_M}} \tilde{\nabla}^2 
\tilde{M}^\prime + \tilde{u}_z \, , \label{GG4}
\end{eqnarray}
%--------------------------------------------------------------------------
with $\tilde{\bf u}=(\tilde{u}_x,\tilde{u}_y,\tilde{u}_z)$ being the velocity field and $\tilde{p}$ the kinematic pressure. The set of equations is closed by an additional expression for the buoyancy field $\tilde{B}(\tilde{\bf x},\tilde{t})$ which follows from (\ref{buoorig}) and takes the dimensionless form of
%--------------------------------------------------------------------------
\begin{equation}
\tilde{B}= -\tilde{z} + \max \left[ \tilde{M}^\prime , \tilde{D}^\prime +\left(1-\frac{Ra_D}{Ra_M} -\mathcal{C}\right) \tilde{z} +\mathcal{S}\right ]. \label{Buoy1}
\end{equation}
%--------------------------------------------------------------------------
The parameters of the system are determined by a unique equilibrium state which is given by the linear profiles $\bar D({z})$, ${\bar M}({z})$, the Brunt-Vaisala frequency $N_s$, the kinematic viscosity $\nu$ and the thermal diffusivity $\kappa$. The dimensionless equations contain five parameters, $Ra_D$, $Ra_M$, $Pr$, ${\cal C}$ and ${\cal S}$. The Prandtl number is given by $Pr=\nu/\kappa$. The dry and moist Rayleigh numbers $Ra_D$ and $Ra_M$ are given by
%--------------------------------------------------------------------------
\begin{equation}
 Ra_D=\frac{({D}_0-{D}_H)H^3}{\nu \kappa} \,, \;\; Ra_M=\frac{({M}_0-{M}_H)H^3}{\nu \kappa}\,. \label{Para1}
\end{equation}
%--------------------------------------------------------------------------
The last two parameters which appear in the saturation condition (\ref{Buoy1}) are related to the phase changes. These are the so-called \emph{surface saturation deficit} $\mathcal{S}$ and the \emph{condensation in saturated ascent} $\mathcal{C}$. The first of both defines how strongly the air is unsaturated at the lower plane. A negative amplitude stands for a deficit of liquid water, a positive for a prescribed amount of liquid water. The second parameter determines how much liquid water can be formed in the convection layer when an air parcel that is saturated at $z=0$ rises adiabatically to 
the top. They are given by
%--------------------------------------------------------------------------
\begin{equation}
\mathcal{S}= \frac{{D}_0 -{M}_0}{{M}_0-{M}_H}\,, \qquad
\mathcal{C}= \frac{N_s^2 H}{{M}_0-{M}_H} \,. \label{Para2}
\end{equation}
%--------------------------------------------------------------------------
Note that the (linear) equations for the dry and moist buoyancy fields are not independent.\cite{Bretherton1,Weidauer1} If we define a scalar
%--------------------------------------------------------------------------
\begin{equation}
\tilde{\Phi}=\tilde{M}^\prime -\frac{Ra_M}{Ra_D}\tilde{D}^\prime
\end{equation}
%--------------------------------------------------------------------------
then we can derive from (\ref{GG3}) and (\ref{GG4}) an advection-diffusion equation for a conserved scalar $\tilde{\Phi}$ which is given by
%--------------------------------------------------------------------------
\begin{equation}
\frac{\partial \tilde{\Phi}}{\partial \tilde{t}} + (\tilde{\bf u} \cdot \tilde{\nabla} ) \tilde{\Phi}= \frac{1}{\sqrt{Pr Ra_M}} \tilde{\nabla}^2 \tilde{\Phi} \, .
\end{equation}
%--------------------------------------------------------------------------
It follows that if $\tilde{M}^\prime$ and $\tilde{D}^\prime$ have different initial conditions they will tend to become equal. Here we use $\tilde{M}^\prime
(\tilde{t}=0)=(Ra_M/Ra_D) \tilde{D}^\prime(\tilde{t}=0)$ as initial condition so that both buoyancy field are synchronized for all times via
%--------------------------------------------------------------------------
\begin{equation}
\tilde{D}^\prime =\frac{Ra_D}{Ra_M} \tilde{M}^\prime \,. \label{couple}
\end{equation}
%--------------------------------------------------------------------------
In fact we only solve one of the Eqns. (\ref{GG3}) and (\ref{GG4}) and obtain the other buoyancy field with the last equation.

The set of equations is solved in a rectangular box of horizontal size $L$ for $x$ and $y$ directions and height $H$ for the $z$ direction. The aspect ratio is $A=L/H$. At the side walls of the domain periodic boundary conditions for all fields are imposed. For the velocity field, we follow the classical works of Kuo and Bretherton and apply free slip BC at both boundary planes. For the buoyancy fields, we distinguish two sets of BC, Dirichlet ($\mathcal{D}$) and Neumann ($\mathcal{N}$) boundary conditions for $D^{\prime}$ and $M^{\prime}$:
%--------------------------------------------------------------------------
\begin{eqnarray}
\mathcal{D} \quad &\Rightarrow& \quad M^{\prime}|_{z=0}=M^{\prime}|_{z=H}=0 \\
\mathcal{N} \quad &\Rightarrow& \quad \frac{\partial M^{\prime}}{\partial z}\Big |_{z=0}=  \frac{\partial M^{\prime}}{\partial z}\Big |_{z=H}=0 \, .
\end{eqnarray}
%--------------------------------------------------------------------------
Liquid water is formed whenever
%--------------------------------------------------------------------------
\begin{equation}
{q}_l({\bf {x}},{t})= {M}({\bf {x}},{t})-[{D}({\bf {x}},{t})-N_s^2 {z}]>0 \, \label{watercon} .
\end{equation}
%--------------------------------------------------------------------------
Therefore we can consider this quantity as being proportional to the liquid water content once it is larger than zero and as a liquid water deficit once it is smaller than zero.

\subsection{Conditionally unstable equilibria}
Convective motion can be triggered out of an equilibrium state in which no fluid motion is present. In our moist convection model different 
equilibrium states are possible which obey different stability properties. The stability of a moist convection layer can then be divided into three categories: absolutely stable, conditionally unstable and linearly unstable (see e.g. Refs. \onlinecite{Emanuel}, \onlinecite{Siebesma}). In an absolutely stable convection layer all air parcels are stable with respect to any vertical displacement and relax back into their initial position. For a linearly unstable environment infinitesimal vertical displacements result in convective motion of unsaturated and saturated air parcels. Both buoyancy fields are unstably stratified. This regime is closest to the original dry Rayleigh-B\'{e}nard convection and has been investigated in Refs. \onlinecite{Olivier1} to \onlinecite{Weidauer1}. It can be considered as an intermediate stage between
the dry convection and the conditionally unstable moist convection.

Of special interest in the atmospheric context is the conditionally unstable equilibrium in which dry air parcels are stably stratified and saturated air parcels are unstably stratified.\cite{Siebesma} The studies in the following are limited to exactly this case. It was studied in Refs. \onlinecite{Weidauer2} and \onlinecite{Olivier2} for the case of Dirichlet boundary conditions. Also the original works by Bretherton\cite{Bretherton1,Bretherton2} are for this case. From Eqns. (\ref{EquiD}) and (\ref{EquiM}) it follows then that ${D}_H>{D}_0$ and ${M}_H<{M}_0$ and thus $Ra_D<0$ and $Ra_M>0$, respectively. One can then distinguish two classes of conditionally unstable equilibrium states (see also the saturation condition Eq. (\ref{buoorig})):
%--------------------------------------------------------------------------
\begin{enumerate}
\item[C1] Equilibrium which is fully unsaturated or fully saturated. This can be achieved, e.g., if  ${M}_0={D}_0$ and thus ${\cal S}=0$.
\item[C2] Equilibrium which is partly saturated and partly unsaturated. The threshold between both determines an {\em equilibrium cloud base} 
which is given by the condition
\begin{equation}
\bar{M}(z_{CB})=\bar{D}(z_{CB})-N_s^2 z_{CB}\,. \label{cloudbase}
\end{equation}
It follows if ${M}_0\ne {D}_0$ and thus ${\cal S}\ne 0$.
\end{enumerate}
%--------------------------------------------------------------------------
More detailed, for the first class C1 one can distinguish thus 3 cases:
%--------------------------------------------------------------------------
\begin{enumerate}
\item[C1a] Fully unsaturated equilibrium with $\bar{M}(z)<\bar{D}(z)-N_s^2 z$ over the whole layer.
\item[C1b] Kuo-Bretherton case with $\bar{M}(z)=\bar{D}(z)-N_s^2 z$ which is neither saturated nor unsaturated over the whole layer.\cite{Kuo,Bretherton1}
\item[C1c] Fully saturated equilibrium with $\bar{M}(z)>\bar{D}(z)-N_s^2 z$ over the whole layer.
\end{enumerate}
%--------------------------------------------------------------------------
As it turns out in the next subsection, the comparison of both sets of boundary conditions translates to compare moist convection that evolved from different initial conditionally unstable equilibrium states.

\subsection{Relation between both sets of boundary conditions}
It is now discussed how the simulations with the two sets of boundary conditions have to be compared. Two strategies are possible, either to run simulations with the same buoyancy fluxes or with the same mean buoyancy values at the planes. Here the second option was chosen. The first strategy turned out to be less feasible due to the transitional properties of the present model.\cite{Weidauer2} The reference point will be thus always the run with $\mathcal{N}$. The corresponding run with $\mathcal{D}$ results from the moist turbulence conditions as they evolve in the former.

\begin{figure}
\includegraphics[scale=0.9]{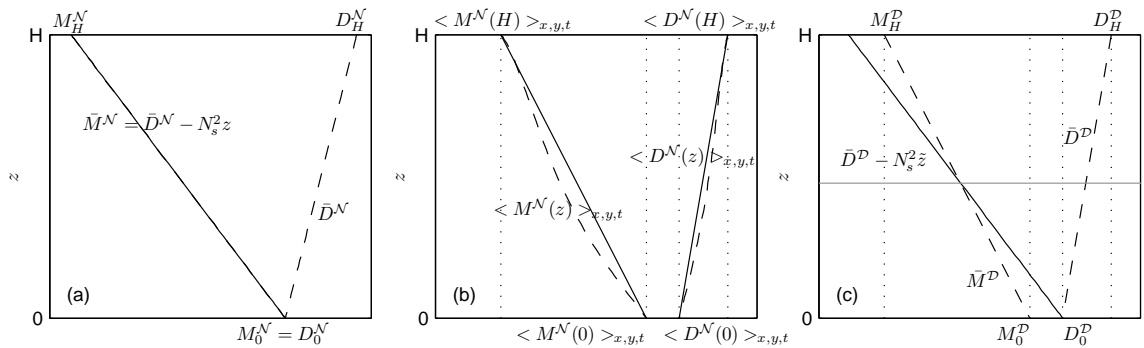}
\caption{Sketch of equilibrium states and their relation. (a) Equilibrium for a run with Neumann boundary conditions ${\cal N}$ in the Kuo-Bretherton regime. The quiescent layer is held therefore exactly at the saturation threshold. (b) Vertical mean buoyancy profiles of ${D}$ and ${M}$ (dashed lines) as obtained with ${\mathcal{N}}$ when moist convective turbulence is generated from case (a). The resulting mean values at the planes determine the linear profiles of the corresponding run with Dirichlet boundary conditions as indicated by thin dotted vertical lines in (b) and (c). (c) Resulting equilibrium for the $\mathcal{D}$ case. Since $N_s$ is unchanged and $\mathcal{S}^{*}\ne 0$ a partly saturated equilibrium results which we denoted as C2. The horizontal line marks the equilibrium cloud base ${z}_{CB}$.}
\label{Equi}
\end{figure}

The comparison is as follows. First, we define a Kuo-Bretherton-type equilibrium state which is case C1b as seen in  Fig. \ref{Equi}(a). The equilibrium state is given by the parameters $\hat{R}a_M,\hat{R}a_D,\hat{\mathcal{S}} (=0)$ and $\hat{\mathcal{C}}$. It is neither subsaturated nor supersaturated. Then we perturb the equilibrium and initiate convective turbulence for the $\mathcal{N}$ case which will evolve into a statistically stationary state. Second, the mean buoyancy profiles for ${D}$ and ${M}$ are calculated. We denote averages over horizontal planes and time by $ \langle\cdot\rangle_{{x},{y},{t}}$ (see Fig. \ref{Equi}(b)). Third, the mean values at the plates determine the equilibrium configuration in the corresponding case $\mathcal{D}$ as shown in Fig. \ref{Equi}(c). The resulting dimensionless parameters $Ra_M^*,Ra_D^*,\mathcal{S}^* (\ne 0)$ and $\mathcal{C}^*$  are now based on these mean values which have to be inserted into (\ref{Para1}) and (\ref{Para2}). From now on
parameter dependencies will be expressed in terms of the second set of dimensionless parameters which are denoted as effective parameters.

Note that a similar translation is necessary in dry Rayleigh-B\'{e}nard convection.\cite{Johnston} At the top plate the resulting mean temperature in the fixed flux case is then smaller than the original temperature of the equilibrium state, at the bottom plate it is higher than the original one. The same is observed in  Fig. \ref{Equi}(a) and (b) of the present setting.
%--------------------------------------------------------------------------
%--------------------------------------------------------------------------

This translation results in an equilibrium for case $\mathcal{D}$ that is partly saturated and partly unsaturated. Unsaturated air is present in the lower part of the layer and saturated air in the upper part (see Fig. \ref{Equi} (c)). Consequently, a cloud base ${z}_{CB}$ is found inside the slab that is given by Eq. (\ref{cloudbase}). Because we start the simulations for the $\mathcal{D}$ and $\mathcal{N}$ case out of different diffusive equilibrium states they have different characteristic quantities that enter the dimensionless parameters. In order to compare both types of runs we rescale the data from the $\mathcal{D}$ case with the characteristic scales of the $\mathcal{N}$ case. For example if the Neumann BC run has a characteristic moist buoyancy difference $\delta{M}^{\mathcal{N}}$ and the Dirichlet BC run has $\delta{M}^{\mathcal{D}}$ then we multiply the buoyancy $B$ of the $\mathcal{D}$ case with $\delta{M}^{\mathcal{D}}/\delta{M}^{\mathcal{N}}$ or in other words with $Ra_M^*/\hat{R}a_M$. Now the buoyancies for both runs are expressed in the same dimensionless units. In this study we want to investigate trends for growing Rayleigh numbers. An increase in the Rayleigh numbers can be performed in two ways: the first way is that we increase the difference ${M}_H-{M}_0$ and leave $\nu$ and $\kappa$ unchanged. The second way is to obtain growing Rayleigh numbers through a decrease of $\nu$ and $\kappa$ such that the Prandtl number remains unchanged. The last way which is chosen here has the technical advantage that velocities and buoyancies are measured in the same units for all runs. The Brunt-Vaisala frequency $N_s$ is also the same.

\subsection{Numerical method}
The equations are solved by a pseudospectral scheme with fast Fourier transformations in the horizontal directions and Chebyshev polynomials in the vertical direction. A 2/3 de-aliasing is used. Parallelization is implemented by dividing the box into vertical slices. Time stepping is done with linear multistep methods. The Courant-Friedrichs-Levy number is always less then 0.3 and the spatial resolution is controlled by the Kolmogorov scale $\eta$. If $\kappa_{min}$ is the smallest resolved length scale then for all simulations $\kappa_{min}\eta \geq 1.5$ holds.
%--------------------------------------------------------------------------
\begin{table}
\begin{ruledtabular}
\begin{tabular}{ccccccc}
$Run$ & $1/\Delta t$ & $N_h$ & $N_z$ & $Ra_M^{*}$ & $\mathcal{S}^{*}$ & $\mathcal{C}^{*}$ \\
\hline
N0/D0 & 100/100 & 128/128 & 17/17 & $5.789 \cdot 10^3$ & 0.61 & 2.69 \\
N1/D1 & 90/90 & 256/256 & 33/33 & $1.427 \cdot 10^4$ & 1.01 & 3.64 \\
N2/D2 & 120/120 & 256/256 & 33/33 & $3.270 \cdot 10^4$ & 1.47 & 4.76 \\
N3/D3 & 150/120 & 512/512 & 65/65 & $1.066 \cdot 10^5$ & 1.45 & 4.87 \\
N4/D4 & 120/150 & 512/512 & 65/65 & $3.530 \cdot 10^5$ & 1.20 & 4.41 \\
N5/D5 & 120/160 & 512/512 & 65/129 & $1.394 \cdot 10^6$ & 0.85 & 3.72 \\
N6/D6 & 200/230 & 1024/1024 & 129/129 & $4.441 \cdot 10^6$ & 0.74 & 3.53 \\
N7/D7 & 200/280 & 1024/1024 & 129/129 & $1.526 \cdot 10^7$ & 0.67 & 3.40 \\
DryN4/DryD4 & 200/200 & 512/512 & 65/65 & $1.254 \cdot 10^5$ & - & -
\end{tabular}
\end{ruledtabular}
\caption{List of parameters of all performed DNS runs. N is for runs with Neumann BC and D for corresponding runs with Dirichlet BC. We always compare runs with the same number, i.e., N2 with D2. $Pr=0.7$ and $A=16$ for all runs.}
\label{table1}
\end{table}
%--------------------------------------------------------------------------

In Table \ref{table1} a list of the performed DNS runs is given with effective physical and numerical parameters. The Prandtl number is $0.7$ for all runs. Without loss of generality we use $\bar{D}(z)=(-1/3) \bar{M}(z)$ throughout the work for the diffusive equilibrium profiles of both buoyancy fields. This is one particular possible choice for a configuration C1b which was always our
starting point. The choice results to $\hat{R}a_D=(-1/3) \hat{R}a_M$. Thus the fluctuations form $M^\prime$ and $D^\prime$ are correlated via (\ref{couple}) and it follows that $Ra_D^*=-1/3 Ra_M^*$. That is why $Ra_D^*$ is not listed separately in Tab. \ref{table1}. The effective parameters are defined in a way that they are the same for both boundary conditions. The diffusive equilibrium for the $\mathcal{D}$ case based on these parameters is of mixed type (see Fig. \ref{Equi} (c)). In addition we run a case for constant fluxes without phase changes denoted as DryN4 and a corresponding run for $\mathcal{D}$ denoted as DryD4. Furthermore run N3 was repeated with aspect ratio 24 to study the impact of different aspect ratios on the dynamics. The number of equally spaced grid points in the horizontal directions is given by $N_h$, the number of the Chebyshev Gauss-Lobatto grid points by $N_z$. They are not equally spaced and tend to cluster at the boundary planes. The time increment is denoted with $\Delta t$.

\section{\label{Abschnitt3}Results}

\subsection{Convection without phase changes}
From now on we will consider dimensionless quantities only and abandon the tilde. Turbulent convection is always triggered out of the equilibrium state, either by an infinitesimal perturbation of the linearly unstable quiescent layer or by a finite-amplitude perturbation of the conditionally unstable layer. Then the system relaxes into a statistically stationary state after passing a transient. Statistical and structural analysis is provided for the statistically stationary regime in the following.

Let us start to compare two runs without phase changes. In order to get to the dry Rayleigh-B\'enard convection equations we can think of $M$ (or also $D$) as being now directly proportional to the temperature $T$. One replaces the buoyancy term $B$ in the momentum equation by $B = M = g\alpha T$ with the gravity acceleration $g$ and the thermal expansion coefficient $\alpha$. The field is again decomposed into a linear and a fluctuating part, $M({\bf x},t)={\bar M}(z) + M^{\prime}({\bf x},t)$.
%--------------------------------------------------------------------------
\begin{figure}
\includegraphics[scale=0.95]{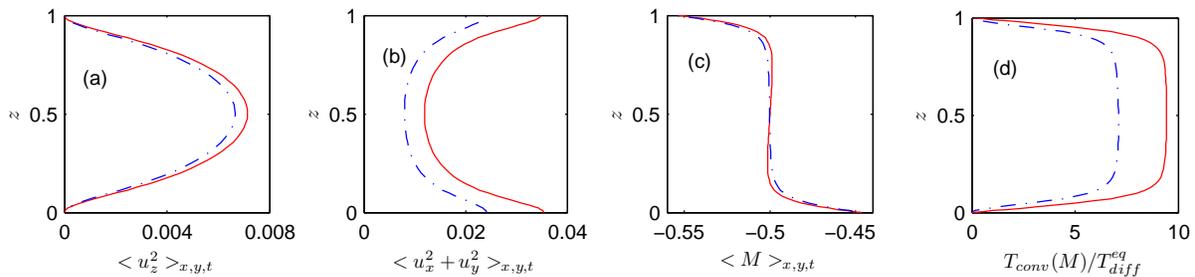}
\caption{(Color online) Mean vertical profiles of  (a) the vertical velocity fluctuations, (b) the horizontal velocity fluctuations, (c) the mean buoyancy (equivalent to temperature) and (d) the convective heat transport. The dash-dotted line represents run DryD4 and the solid line run DryN4.}
\label{Dry}
\end{figure}
%--------------------------------------------------------------------------

In Fig. \ref{Dry} we plot mean profiles of some quantities with respect to $z$ in order to compare them later with the results of the runs with phase changes. Symmetry with respect to the mid plane is an intrinsic property of dry Rayleigh-B\'{e}nard convection and established here. Velocity fluctuations in all three directions (see panels (a) and (b)) are enhanced in the constant flux case $\mathcal{N}$ in comparison to the constant amplitude case $\mathcal{D}$. The mean buoyancy (or temperature) profiles are compared in panel (c). The typical boundary layers at the planes form which enclose a well-mixed bulk. The convective buoyancy flux in panel (d) is also enhanced for $\mathcal{N}$. Both, the mean buoyancy (or temperature) and the vertical velocity fluctuation profiles for the $\mathcal{D}$ case are qualitatively similar to what was found in DNS of dry convection with large aspect ratio such as in Ref. \onlinecite{Hardenberg}, although stress-free instead of no-slip BC for the velocity field are used here. The reason for the higher amplitudes in case $\mathcal{N}$ might be due to the flow patterns that we observe. For the case $\mathcal{N}$, a large diamond-like structure of the buoyancy field $B$ is found in combination with a ribbon of downwelling fluid motion (see also later in the text) is observed. Again, this could be a relic of the fact that the critical wavenumber for the onset of convection is zero.\cite{Chapman} Such a pronounced flow structure enhances the fluctuations.

In large-eddy simulations of dry convection in large-aspect-ratio cells, Fiedler and Khairoutdinov \cite{Fiedler} found similar extended structures in the buoyancy field in combination with smaller patterns of the velocity field. These simulations are close to the present case ${\mathcal N}$ in terms of the BC. We therefore conclude that the applied Smagorinsky sub grid-scale model results in effective Rayleigh numbers that are comparable with our values. This might be the reason why we observe similar structures. This observation for moderate Rayleigh numbers is also in line with the findings of Johnston and Doering.\cite{Johnston}

Of particular interest in turbulent convection are always the vertical transport properties. The global transport is measured by a dimensionless Nusselt number $Nu$. The turbulent heat transport is composed of a convective $T_{conv}$ and a diffusive flux $T_{diff}$ and normalized by the diffusive flux at equilibrium $T_{diff}^{eq}$. Here the mean values of the buoyancy fields at the top and bottom planes are known in both cases due to the boundary conditions. Thus the mean gradient of the moist buoyancy in dimensional form is given by $({M}_0^{\mathcal{D}}-{M}_H^{\mathcal{D}})/H$.

As stated above, in convection without phase changes the dry and moist buoyancies are equivalent. Since it makes for this section no difference we will already base the definition of the Nusselt number, as it will be used for the rest of the work, on the moist buoyancy diffusive flux of case $\mathcal{D}$. The Nusselt number with respect to any of the three buoyancy fields $A=D, M$ or $B$ is thus defined by
%--------------------------------------------------------------------------
\begin{equation}
Nu_A(z)=\frac{T_{conv}(A)+T_{diff}(A)}{T_{diff}^{eq}} =\frac{H\left(\langle u_zA\rangle_{x,y,t}-\kappa\frac{\partial \langle A
\rangle_{x,y,t}}{\partial z}\right)}
{\kappa(M_0^{\mathcal{D}}-M_H^{\mathcal{D}})}\,.
\label{Nusselt1}
\end{equation}
%--------------------------------------------------------------------------
The Nusselt numbers for both dry runs are $Nu^{\mathcal{N}}=9.32$ and $Nu^{\mathcal{D}}=7.07$ (see Table \ref{table1} for the parameters). Recall that there exists no dependence of $Nu$ on the vertical coordinate $z$ in dry convection. This is a direct consequence of averaging the buoyancy equation in planes at fixed height and in time. At this point we wish to state also that for the moist convection case other definitions are possible, e.g. by relation to the dry buoyancy flux.\cite{Schumacher1}

\subsection{Convection with phase changes}

\subsubsection{Mean buoyancy profiles and cloud cover\label{Gardine}}
\begin{figure}
\includegraphics[scale=0.95]{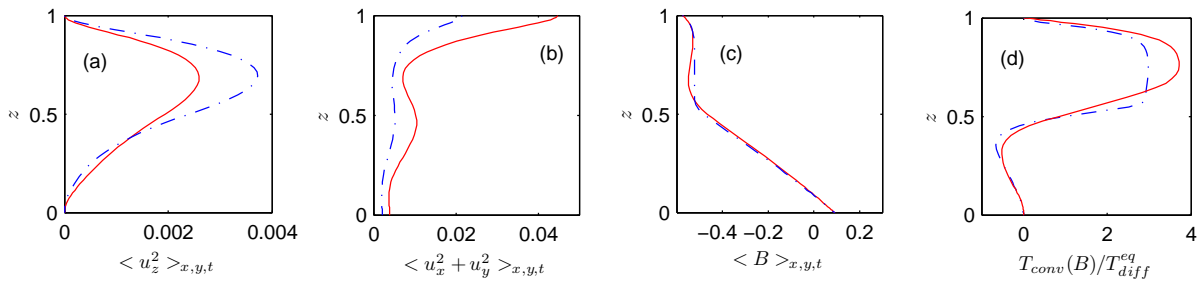}
\caption{(Color online) Mean vertical profiles of  (a) the vertical velocity fluctuations, (b) the horizontal velocity fluctuations, (c) the mean buoyancy and (d) the convective buoyancy transport.  The dash-dotted line represents run D4 and the solid line run N4.}
\label{compare}
\end{figure}

We turn now to the conditionally unstable moist convection case. In order to emphasize the differences of Rayleigh-B\'enard convection with and without phase 
changes the same quantities are shown in Fig. \ref{compare} as in Fig. \ref{Dry} for the dry convection case.  In panels (c) and (d) we display the buoyancy $B$ 
and the corresponding convective flux $T_{conv}(B)$. The main difference is the breaking of the top-down symmetry of all vertical profiles, a consequence 
of the latent heat release that is a source of additional kinetic energy in the upper part of the convection domain. As discussed further below, the convective 
buoyancy transport can now become even negative since the lower part of the layer is still stably stratified.
%--------------------------------------------------------------------------
%--------------------------------------------------------------------------
\begin{figure}
\includegraphics[scale=0.9]{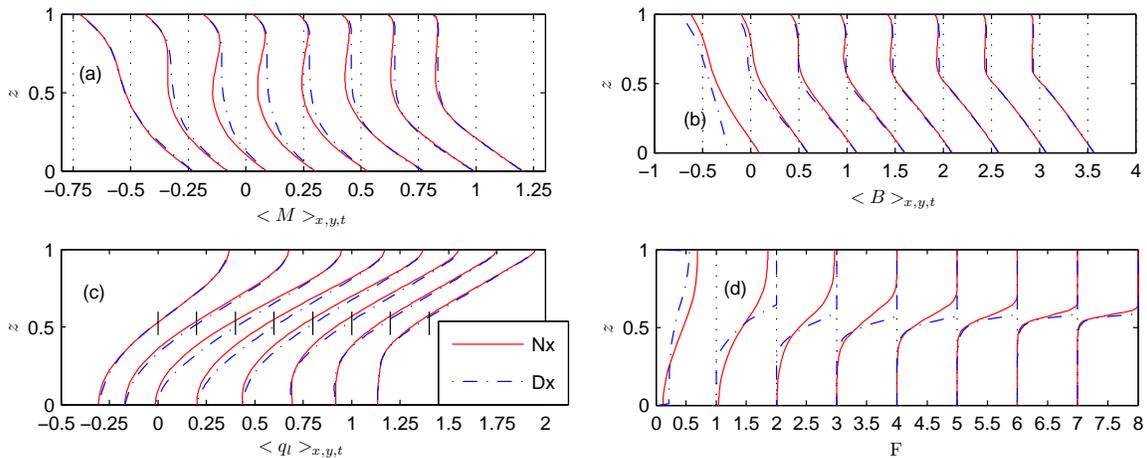}
\caption{(Color online) Vertical dependence of various mean quantities. The data for the different runs (N0--N7 and D0--D7) are successively shifted. The Rayleigh number increases from left (run 0) to right (run7). (a) Moist buoyancy profiles which are shifted by 0.2. (b) Buoyancy profiles $B$ using eq. (\ref{Buoy1}) which are shifted by 0.5. (c) Liquid water content profiles which are shifted by 0.2. (d) Cloud fraction profiles following from Eq. (\ref{cloud}) which are shifted by 1. Line style is the same in all four panels.}
\label{Bild4}
\end{figure}
%---------------------------------------------------------------------------------

\begin{figure}
\includegraphics[scale=1.2]{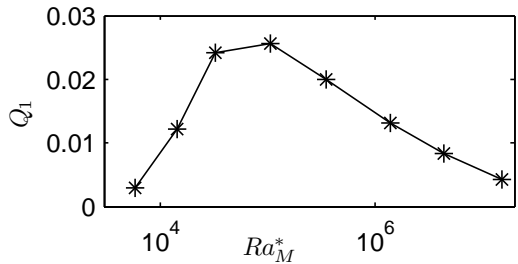}
\caption{Difference of the vertical moist buoyancy profiles for corresponding runs of cases $\mathcal{D}$ and $\mathcal{N}$ using Eq. (\ref{Q1}). Data are plotted against the effective moist Rayleigh number.}
\label{Bild5}
\end{figure}

Let us now discuss the results of the moist convection runs. As mentioned in the introduction, we will monitor the turbulence properties as a function of the Rayleigh number and compare both BCs. We start with the vertical mean profiles of quantities that are directly related to $ M^\prime$. The first inspection of the vertical profiles in Fig. \ref{Bild4} unravels a similar asymmetry which we already mentioned in Fig. \ref{compare} and which is discussed in detail in previous studies.\cite{Bretherton2,Olivier1,Schumacher1} In Fig. \ref{Bild4}(a) it is demonstrated that the mean moist buoyancy coincides for both cases at the boundaries. The profiles become increasingly asymmetric with increasing Rayleigh number. It is very close to the linear diffusive equilibrium profile in the lower part of the layer which is a hint for 
almost absent turbulent mixing -- consequence of the stable stratification. The profiles turn to be constant with height in the upper part. This in turn is a clear indication for well-mixed turbulence with a small boundary layer at the top. A further interesting result is that $\langle M^{\mathcal{D}}(z)\rangle_{x,y,t}>\langle M^{\mathcal{N}}(z)\rangle_{x,y,t}$. Since the moist buoyancy is relevant for many other quantities a measure is defined that quantifies how much the profiles differ for the two cases of boundary conditions,
%---------------------------------------------------------------------------------
\begin{equation}
Q_1=\Big\langle|\langle M^{\mathcal{N}}\rangle_{x,y,t}- \langle M^{\mathcal{D}}\rangle_{x,y,t}|\Big\rangle_{z} \,.
\label{Q1}
\end{equation}
%---------------------------------------------------------------------------------
A decrease of $Q_1$ for $Ra_M^* \gtrsim 10^5$ is observable as shown in Fig. \ref{Bild5}. The quantity indicates that the differences diminish progressively. In Fig. \ref{Bild4}(b) the mean profiles of the buoyancy $B$ follow. Again, one can detect a difference for the lower and upper parts of the layer which is similar to that of the moist buoyancy $M$. For the highest Rayleigh numbers the mean buoyancy profiles become almost indistinguishable.
%---------------------------------------------------------------------------------
%---------------------------------------------------------------------------------

We know from (\ref{couple}) and (\ref{watercon}) that $M$ is also directly related to the water content. In Fig. \ref{Bild4}(c) the mean profiles of $q_l$ are shown. The saturation threshold for each data set is marked as a vertical solid line segment. Similar to the mean profiles of $M$, the profiles for the water content converge to each other with increasing Rayleigh number. The fixed amplitude runs obey always slightly more liquid water across the domain than the constant flux runs, except for the run 6 in the lower part of the box. A possible explanation could be that the fixed amplitude case sustains the water content to a constant value throughout the whole DNS run. In the constant flux case, this water content can fluctuate in space and time across the boundary planes and thus affect the cloudiness across the whole layer.

Furthermore, we show in Fig. \ref{Bild4}(d) the cloud fraction as a function of $z$ which is defined by ($\mathbf{I}$ is the indicator function)
%---------------------------------------------------------------------------------
\begin{equation}
F(z)=\langle\mathbf{I}_{q_l>0}(z)\rangle_{x,y,t} \, . \label{cloud}
\end{equation}
%---------------------------------------------------------------------------------
For the runs D0 and N0--N2 one can observe subsaturated air at all heights which corresponds with an open cloud layer. All other runs obey a closed cloud layer in the upper and completely unsaturated air in the lower part. In between subsaturated and saturated air coexist and the percentage of saturated air increases with height. The cloud cover profiles converge again such that the 
albedo for different thermal boundary conditions would be nearly the same within our MRBC model.

The thickness of this intermediate layer is different for the two boundary conditions and can be explained as follows: the liquid water content is slightly higher for the $\mathcal{D}$ case as shown in panel (c) which should give a bigger cloud fraction. For the case $\mathcal{N}$ we found that the buoyancy $M^\prime$ has a higher variance about the mean value (not shown). This results in a higher variation of the liquid water content and thus a thicker crossover layer from a completely cloudless state to a full cloudy one. The thickness of this intermediate layer and the difference in the thickness for the two boundary conditions are also decreasing with increasing Rayleigh number, respectively. It is known that for higher Rayleigh numbers the individual plumes become more fragmented. Furthermore diffusion decreases  and the degree of stable stratification in the lower part increases. The stronger fragmented moist filaments penetrate less easily into the dry layer and evaporate. This all seems to result in a thinner transition layer between both regions.

\subsubsection{Buoyancy transport\label{Gurke}}
The vertical transport properties in the conditionally unstable parameter regime are investigated in the following. From Eq. (\ref{Nusselt1}) we adapt the following definitions
%--------------------------------------------------------------------------------
\begin{eqnarray}
Nu_A^{\mathcal{N}}(z)&=&\left[ \sqrt{Pr\hat{R}a_M} \langle u_z A\rangle_{x,y,t}-\frac{\partial \langle A\rangle_{x,y,t}} {\partial z} \right] \frac{\hat{R}a_M}{Ra_M^{*}}\,, 
\nonumber \\
Nu_A^{\mathcal{D}}(z)&=&\left[\sqrt{PrRa_M^{*}} \langle u_z  A\rangle_{x,y,t} -\frac{\partial\langle A\rangle_{x,y,t}} {\partial z} \right] \label{NuNeumann} \,,
\end{eqnarray}
%--------------------------------------------------------------------------------
\begin{figure}
\includegraphics[scale=0.9]{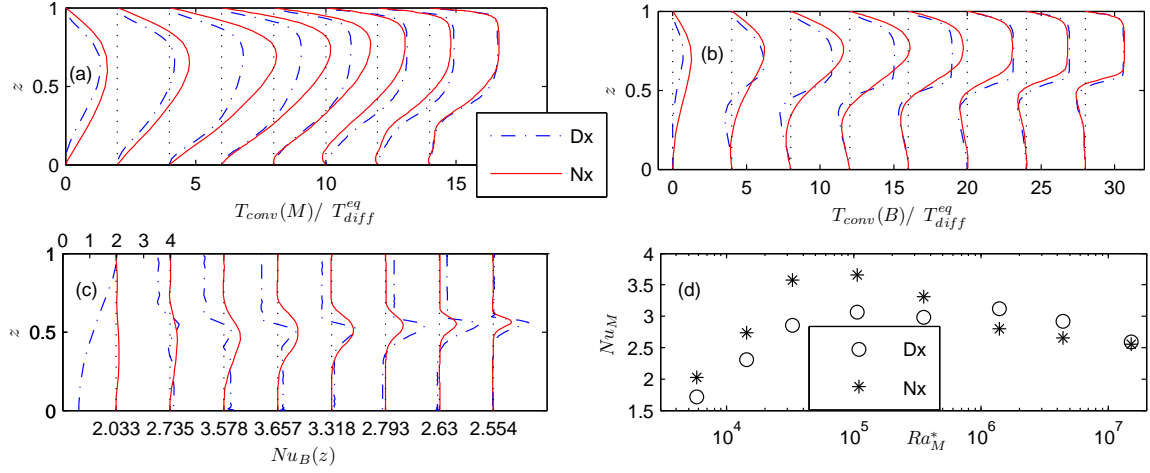}
\caption{(Color online) Vertical profiles of mean buoyancy transport (from left to right for runs 0 to run 7). The profiles for the different runs are again successively shifted. (a) Profiles of the convective transport of the moist buoyancy $T_{conv}(M)/T_{diff}^{eq}$ which are shifted by 2. (b) Profiles of the convective buoyancy flux  $T_{conv}(B)/T_{diff}^{eq}$ which are shifted by 4. (c) Profiles of the Nusselt number $Nu_B$ which are shifted by 2 as indicated at the top of the panel. Here the curves are centered about the Nusselt number for $B$ for the $\mathcal{N}$ case at height zero. This value is always indicated at the bottom and by a vertical dotted line. Same line style for (a)-(c). (d) Nusselt number of the moist buoyancy plotted against the effective moist Rayleigh number $Ra_M^*$.}
\label{Bild6}
\end{figure}
%--------------------------------------------------------------------------------
where $A=D, M$ or $B$.

Note that $Nu_D(z)$ and $Nu_M(z)$ for both BC are still constant functions as in dry convection. Only the Nusselt number with respect to the buoyancy $B$ will violate this property since it incorporates the impact of phase changes. We focus to the convective transport and discuss $T_{conv}(M)$ in Fig. \ref{Bild6}(a). It was verified that the diffusive contribution has exactly the same shape, is non-zero at $z=0$ and negative such that both sum up to a constant non-zero positive constant as given by (\ref{Nusselt1}) or (\ref{NuNeumann}). In the upper part a fully developed turbulent layer emerges and the convective transport of $M$ takes the maximum value. In the lower part of the layer it becomes almost zero with increasing Rayleigh number. For runs N5--N7 close to the bottom plane a layer of weak negative convective transport emerges. It is most strongly in N5 and decreases for N6 and N7. This behavior might arise due to falling plume structures which are a result of the possible variations of $D$ and $M$ in the 
boundary plane together with mean flow structures which we will discuss in Sec. III.B.4.

Additionally we plot the transport properties of the resulting buoyancy $B$ in Figs. \ref{Bild6}(b) and (c). In panel (b) of this figure the convective transport $T_{conv}(B)$ is shown. The upper part of the curves is similar to those of $T_{conv}(M)$, a result of (\ref{Buoy1}). However, the lower part of the curves shows negative values since the buoyancy $B$ is related to the dry buoyancy $D$ as determined by (\ref{couple}). This is a manifestation of the stable stratification of the lower part of the convection layer. Upward directed velocity and positive buoyancy fluctuation are anti-correlated.  As we increase the Rayleigh number the transport in the lower part of the layer, which has to rely solely on diffusion, becomes increasingly inefficient for both sets of boundary conditions. A destabilization requires an extension of our MRBC model, e.g. by a radiative bulk cooling as recent complementary studies showed. \cite{Pauluis3} The profile of the convective transport for $z\ge 0.5$ in this case (see runs N7 and D7 in Fig. 6 (b)) is qualitatively equal to the one in Fig. 2(b) for classical dry Rayleigh-B\'{e}nard convection. Here convective transport is fully established. The turbulent motion in the dry convection layer is strongly suppressed, in the cloudy layer above it is amplified.

In panel (c) of Fig. \ref{Bild6}, $Nu_B(z)$ is given for both BC. The Nusselt number of $B$ is not any more necessarily constant with respect to $z$. The profiles are centered about  the value $Nu_B(z=0)$ of the case $\mathcal{N}$. In the zone where unsaturated and saturated air coexist the Nusselt number has a bump which is a result of the latent heat release due to permanently ongoing phase changes. In the upper and lower parts of the layer the profiles are nearly perfectly constant about the height since the air is either fully unsaturated or fully saturated.

The Nusselt numbers of $M$ for both BC are compared in Fig. \ref{Bild6}(d). Note that the values for the corresponding $\mathcal{N}$ are fixed through the boundary condition and are given by definition as $\hat{R}a_M/{R}a_M^*$. For both cases the convective transport decreases in the dry lower layer compared to the diffusive transport with increasing Rayleigh number. The difference between both cases is decreasing as seen from the profiles. This indicates that the vertical transport of heat and moisture becomes almost the same for the highest Rayleigh numbers accessible. It can also be seen that the buoyancy transport in the conditionally unstable regime is rather slightly decreasing or almost constant with $Ra_M^*$ than increasing as known from dry convection. In Ref. \onlinecite{Olivier2}, an upper bound was predicted which is a consequence of the dominant and stably stratified layer of unsaturated air above the bottom plane.

For imposed fluxes the Nusselt number at the bottom and top plane is prescribed by the BC. In Fig. \ref{Bild6} one can see a qualitative change between runs D4 and D5. For example, in Fig. \ref{Bild6}(c) one observes that in runs D1 to D4 the Nusselt number $Nu_B$ is smaller at the upper and larger at the lower boundary than in  runs N1 to N4. The opposite is seen for runs 5 and 6. In run 7 almost no difference at the plates is noticeable. The origin of these differences is in the diffusive equilibrium states. As listed in Tab. \ref{table1},  we can see that both parameters, $\mathcal{S}^*$ and $\mathcal{C}^*$, pass a local maxima at runs 3 and 4 and decrease for higher Rayleigh numbers. Thus less liquid water can be formed which rationalizes the observations mentioned above.

\subsubsection{Mean velocity fluctuations}
%--------------------------------------------------------------------------
\begin{figure}
\includegraphics{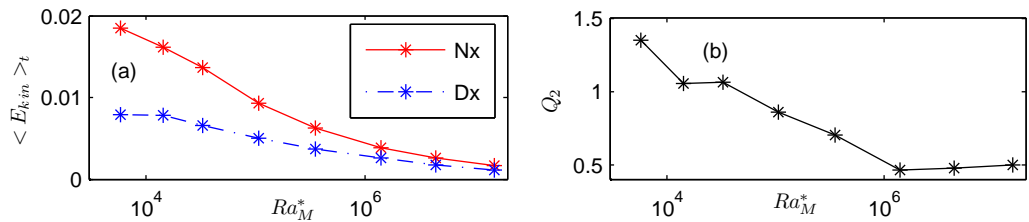}
\caption{(Color online) (a) Time average of the kinetic energy versus the moist effective Rayleigh number. (b) The relative deviation for the kinetic energies using Eq. (\ref{Lupe})
versus $Ra_M^*$.}
\label{Bild7}
\end{figure}
%--------------------------------------------------------------------------------
In the following, we list our findings for the velocity fluctuations. Fig. \ref{Bild7}(a) compares the turbulent kinetic energy (TKE) for all runs, which is given by the time average of
%--------------------------------------------------------------------------------
\begin{equation}
E_{kin}(t)=\frac{1}{2}\langle {\bf u}^2\rangle_{x,y,z}\, .
\end{equation}
%--------------------------------------------------------------------------------
The TKE values for both sets of BC converge to each other with growing Rayleigh number. The magnitude of the TKE in cases $\mathcal{N}$ is by a factor of 1.4 to 0.5 larger than in the corresponding case $\mathcal{D}$. The relative deviation of both is quantified by
%--------------------------------------------------------------------------------
\begin{equation}
Q_2=\frac{\langle E_{kin}^{\mathcal{N}}\rangle_t - \langle E_{kin}^{\mathcal{D}}\rangle_t}{\langle E_{kin}^{\mathcal{D}}\rangle_t} \, . \label{Lupe}
\end{equation}
%--------------------------------------------------------------------------------
Two points should be mentioned to understand the results for the TKE. First it is important to realize that for conditionally unstable settings all the kinetic energy has its origin in latent heat release. In particular for the runs at higher Rayleigh numbers latent heat is released in the upper part of the box only where cloudy air is present. This upper layer is the domain where convective motion is driven only. Second, the dry layer in the lower part is stably stratified and tends to suppress turbulent fluid motion as discussed above. We already observed in the last subsections that the differences between both series decrease for the mean profiles of several physical quantities. The same is now observed for the measure $Q_2$ as can be seen in Fig. \ref{Bild7}(b).
%--------------------------------------------------------------------------------
\begin{figure}
\begin{center}
\includegraphics[scale=0.9]{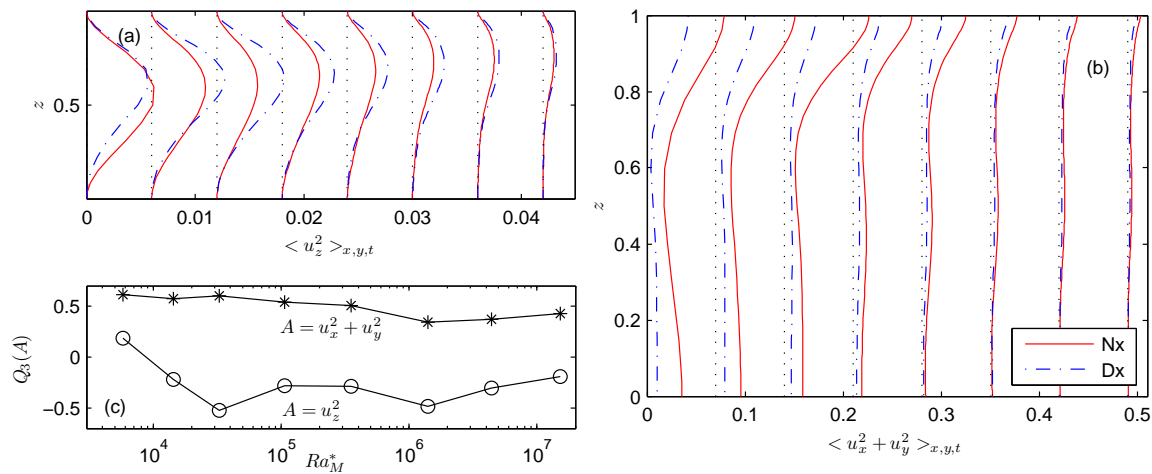}
\caption{(Color online) (a) Vertical mean profiles of the vertical velocity fluctuations. We compare both BC from left to right in runs 0 to 7 which are successively shifted by 0.006. The dotted vertical lines mark zero for each pair of runs (same in panel (b)). (b) Vertical mean profiles of the horizontal velocity fluctuations here shifted by 0.07. (c) Shown is quantity $Q_3$ as defined in (\ref{Q3}). Line style is the same for (a) and (b).}
\label{Bild8}
\end{center}
\end{figure}
%--------------------------------------------------------------------------

In order to detect where the differences in TKE are resulting from, we refine the analysis and plot in Fig. \ref{Bild8} the horizontal and vertical mean square velocity fluctuations as a function of height. In particular, we add always the fluctuations of the two horizontal velocity components. It is seen that the larger fraction of the kinetic energy is always contained in the horizontal velocity components $u_x$ and $u_y$, even when dividing their joint contribution by two. In panel (a) of Fig. \ref{Bild8}, it is observed that with increasing Rayleigh number the vertical velocity fluctuations tend to ever smaller magnitudes and are almost zero. This is in agreement with recent studies in Ref. \onlinecite{Schumacher1} and \onlinecite{Olivier2}. In cloudy regions the fluctuations are enhanced due to the latent heat release, but they decrease with increasing Rayleigh number since plumes which drive the turbulent flow become increasingly fragmented. The relative difference between both series of runs is shown in Fig. \ref{Bild8}(c) as the lower curve. It is quantified by
%--------------------------------------------------------------------------------
\begin{equation}
Q_3(A)=\frac{\langle A^\mathcal{N}\rangle_{x,y,z,t}-\langle A^\mathcal{D}\rangle_{x,y,z,t}}{\langle A^\mathcal{N}\rangle_{x,y,z,t}} \, , \label{Q3}
\end{equation}
%--------------------------------------------------------------------------------
with $A$ either $u_x^2 + u_y^2$ or $u_z^2$. The vertical velocity fluctuations in the case of Dirichlet BC are larger than those of the case of Neumann BC. This results in a negative ratio $Q_3$ for all runs except run 0.

The profiles of the horizontal velocity fluctuations are shown in Fig. \ref{Bild8}(b). For low Rayleigh numbers it can be seen that the fluctuations are enhanced towards both boundary planes in comparison to the center of the layer, mostly due to rising and falling air hitting the boundaries. With growing Rayleigh number the fluctuations decrease everywhere in the layer. The lower part gets increasingly stably stratified and thus diminishes the global convection motion in the whole layer. Falling plumes from the upper layer are then strongly decelerated. The corresponding fluid motion is redirected horizontally. The fluctuation profiles are nearly homogeneous below the cloud base. In contrast to the vertical velocity fluctuations, Fig. \ref{Bild8}(b) shows now that the kinetic energy in the horizontal velocity components is higher for the case of Neumann BC than for Dirichlet BC. This causes a positive $Q_3$ for all runs.
%--------------------------------------------------------------------------------
\begin{figure}
\begin{center}
\includegraphics[scale=0.86,trim=0 85mm 0 0]{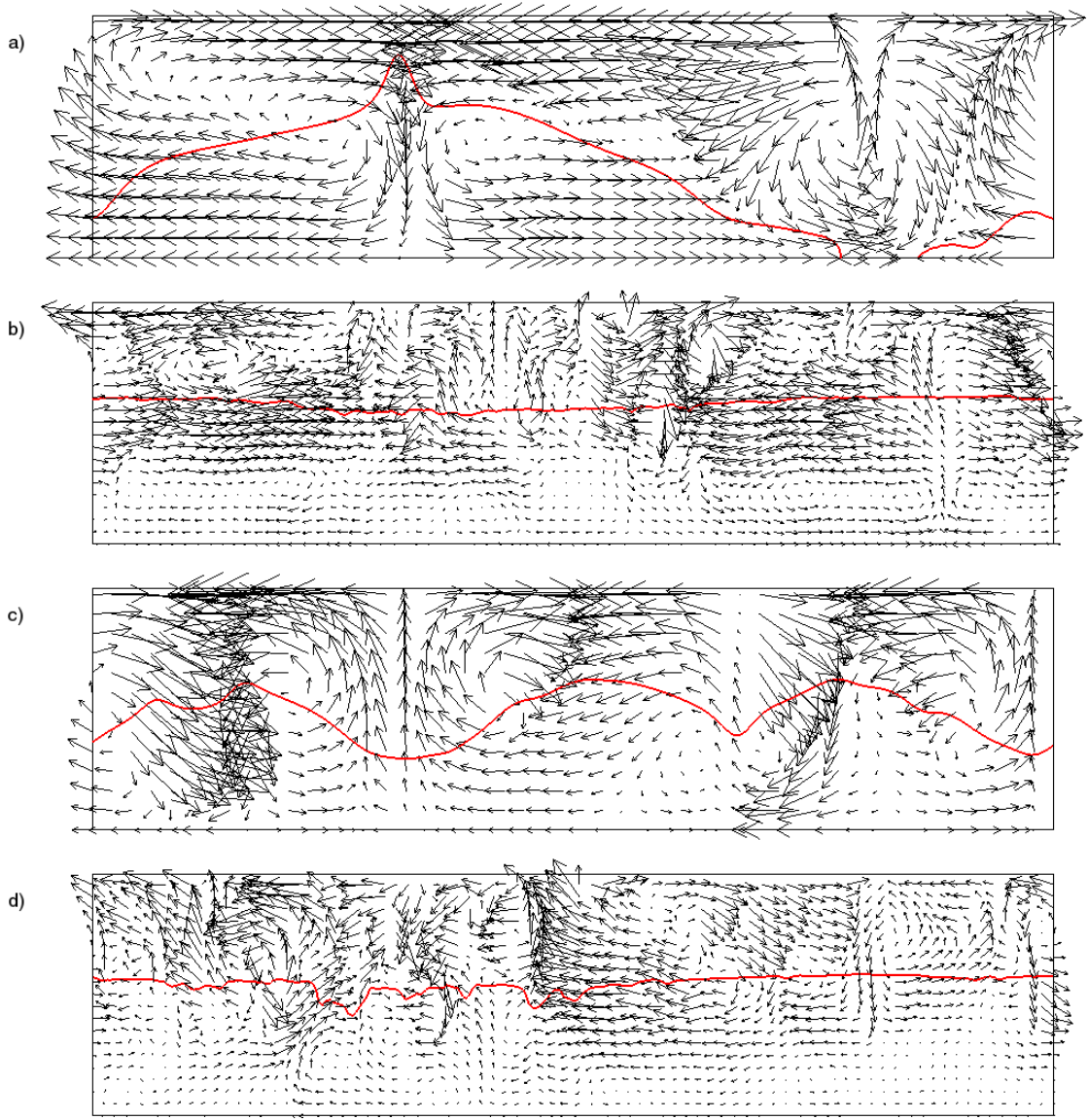}
\caption{(Color online) Vector plots of the instantaneous velocity field. Two-dimensional projections into $x-z$ cross-sections of the simulation domain are displayed. The solid line indicates the cloud boundary. (a) N1, (b) D1, (c) N7 and (d) D7. In order to resolve the vertical fine flow structures better, the size of the large-aspect-ratio domain and the velocity vectors are rescaled.}
\label{Bild14}
\end{center}
\end{figure}
%--------------------------------------------------------------------------

In order to get a more complete picture we add in Fig. \ref{Bild14}  two-dimensional projections of the velocity field in $x-z$ planes for runs at low and high Rayleigh numbers. For runs N1 and D1 in panels (a) and (c), one can observe that up- and downwelling convective motion fills the whole layer. The cloud boundary fluctuates correspondingly over the full vertical extension. The runs N7 and D7 show a different picture. The Rayleigh number is significantly larger and thus the vertical diffusive transport in the stably stratified environment significantly smaller. The two figures (b) and (d)  show also that the up- and downwelling flow between the two layers on top of each other is reduced. The system can be considered as being composed of two increasingly independent subsystems which are increasingly weaker coupled across an ever smaller inversion layer in which the cloud base fluctuates.

\subsubsection{Flow, cloud and buoyancy field structure}
%--------------------------------------------------------------------------
\begin{figure}
\begin{center}
\includegraphics{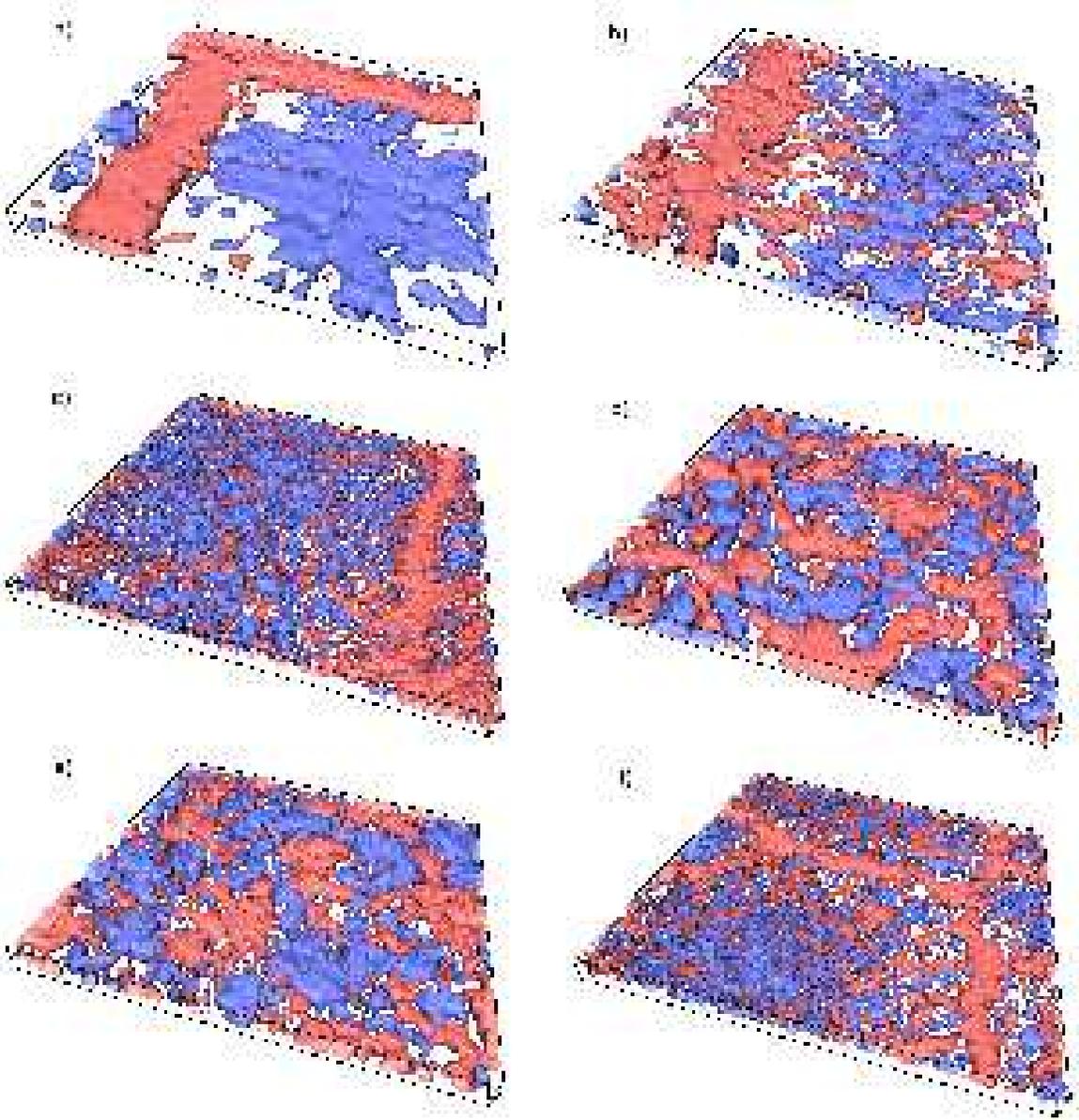}
\caption{Isosurfaces of the mean vertical velocity $\langle u_z\rangle_t$ taken at two levels, red is for 10 per cent of the maximal downward velocity, blue for 10 per cent of the maximal upward velocity. (a) run N2, (b) run N3, (c) run N7, (d) run D2, (e) run D3 and (f) run D7. Aspect ratio is 16 for all cases and the view is from above onto the layer.}
\label{Bild10}
\end{center}
\end{figure}
%--------------------------------------------------------------------------------
As stated in the last section we will shed more light on the flow and buoyancy field structures in the following. Fig. \ref{Bild10} shows isosurfaces of the upward and downward velocity in the layer. The data are the result of a time-averaging over a sequence of at least 40 snapshots which were written out equidistantly for a time lag of 400-800 free-fall time units $T_f$. It is seen that the imposed flux runs yield much more pronounced large-scale flow patterns than the fixed amplitude cases, in particular at the smaller Rayleigh numbers. For example, a roll-like region of downwelling flow can be observed in panels (a) and (b), but not in the corresponding cases shown in panels (d) and (e). A larger aspect ratio does not affect such roll-like structures which we confirmed in a run similar to run N3 with aspect ratio of 24 instead of 16. Similar flow structures can also be observed indirectly via the cloud pattern in nature.\cite{Atkinson} With increasing Rayleigh number the mean flow patterns develop more fine-scale features. It is also known from previous studies \cite{Olivier2} that in case of Dirichlet BC  cloud aggregates increase with growing Rayleigh number. In Fig. \ref{Bild10}(d) we observe a number of up- and downwelling areas. With increasing Rayleigh number these patches merge to bigger patches of dominantly up-or downwelling flow motion.  For the runs D5, D6 and D7 we observed eventually two regions of upward motion only, a major larger one and a smaller secondary region separated by a ribbon of downward moving air. Cases N7 and D7 appear much more similar in their mean flow structure than the runs at lower Rayleigh numbers as the comparison of panels (c) and (f) demonstrates.

On the basis of the evolution of the flow structures we can rationalize why $Q_3$ is negative for the vertical velocity fluctuations and positive for the horizontal fluctuations as seen in Fig. \ref{Bild8}. In the Neumann case we see that the mean flow appears in form of extended roll structures, for the Dirichlet case in form of convection cells. This results in higher vertical velocity fluctuations for the Dirichlet case compared to the Neumann case. In case of the horizontal velocity fluctuations the situation is opposite.  Larger velocity fluctuation amplitudes are observed in the Neumann case in connection
with the roll pattern. For the highest Rayleigh numbers in runs 5 to 7, these mean structures persist and therefore both measures,  $Q_2$ and $Q_3$, remain nearly unchanged. We see in panel (f) of Fig. \ref{Bild10} that extended convection cells in the Dirichlet case disappeared as they were observable in panel (d) of the same figure.
%--------------------------------------------------------------------------
\begin{figure}
\includegraphics[scale=0.65]{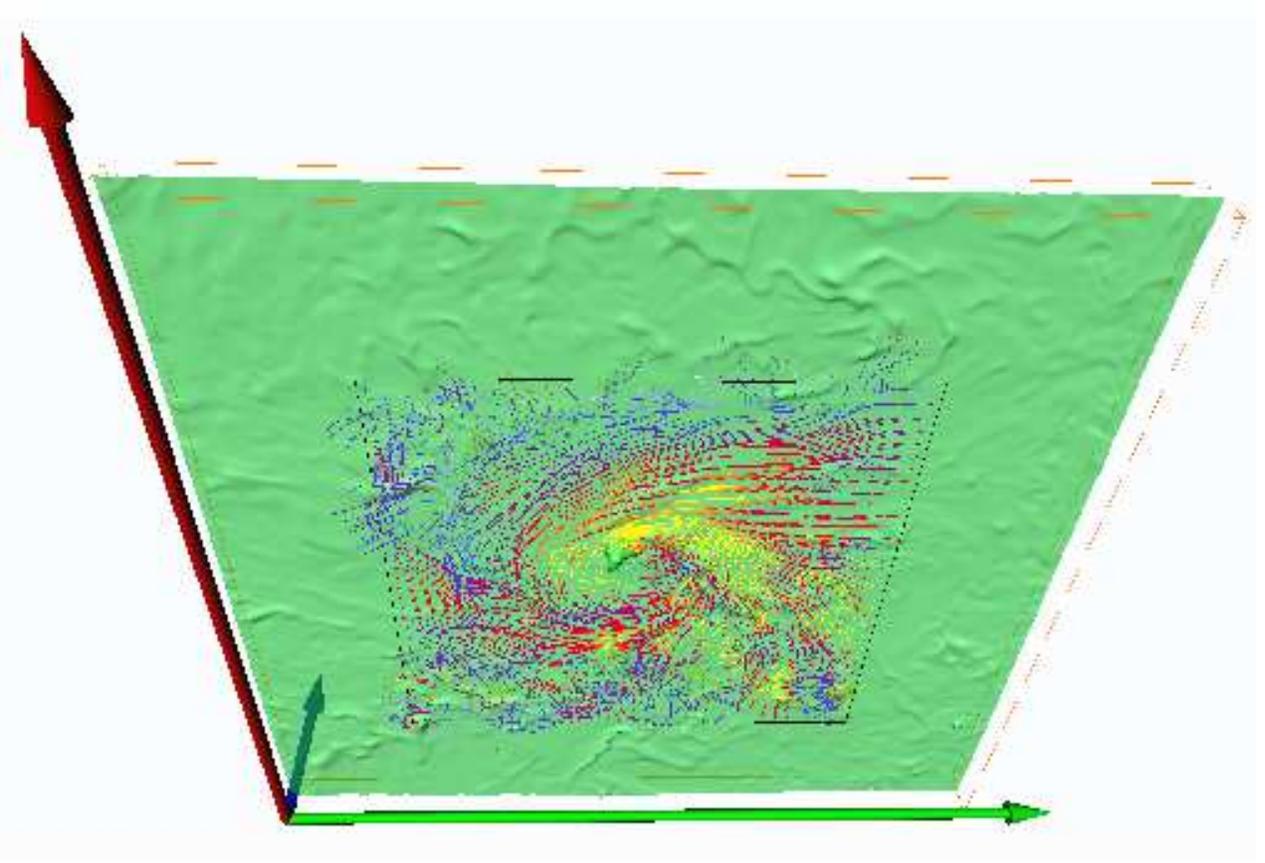}
\caption{Example of a vortex as found in run N6. The look is from below into the box. The green isosurface is the cloud boundary with $q_l=0$. Also shown are streamline segments seeded in a plane at height $z=0.36$ below the cloud boundary. }
\label{Tomate}
\end{figure}
%--------------------------------------------------------------------------------

In most simulations the velocity field is characterized by falling or rising plumes. An exception are the runs N5 to N7 as mentioned already in Sec. \ref{Gurke}. They are in addition accompanied by large vortices as shown in Fig. \ref{Tomate}. The figure displays nicely the correlations between the upwelling and downwelling fluid and the existence of cloud aggregates. In the core of this vortex we find highly buoyant upwelling air which is correlated to a deepening of cloud boundary in Fig. \ref{Tomate}.  The formation of the vortices might be a result of the stable stratification which however does not prevent horizontal fluctuations of the flow triggered by the buoyancy fields which can fluctuate down to the boundary plane in the Neumann case.
%--------------------------------------------------------------------------------
\begin{figure}
\includegraphics{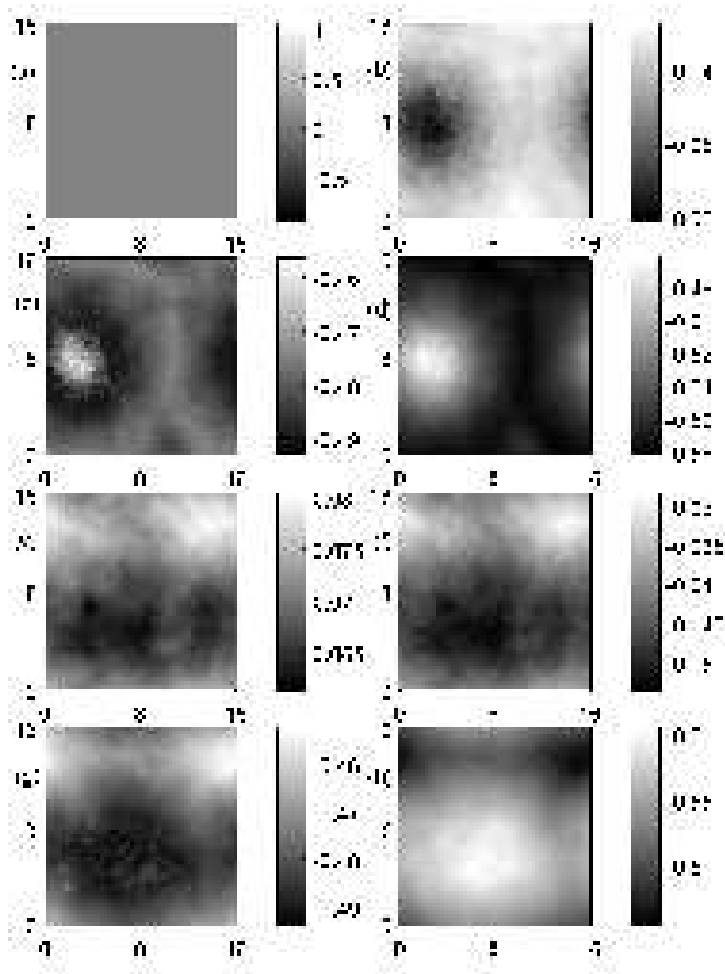}
\caption{Time averaged buoyancy for different heights $z_i$. The two upper rows is for run D6, the two lower rows for the corresponding run N6. 
(a) and (e) $z_1$=0, (b) and (f) $z_2$=0.1135, (c) and (g) $z_3$=0.5, (d) and (h) $z_4$=0.8865. Panel (a) has to be constant due to the boundary 
conditions in cases $\mathcal{D}$.}
\label{Bild12}
\end{figure}
%--------------------------------------------------------------------------------

The time-averaged buoyancy field $\langle B(x,y,z_i)\rangle_t$ is shown in Fig. \ref{Bild12} at four different heights $z_i$ to connect the mean flow structures we showed above to this quantity. We compare runs D6 in the upper four panels of the figure with run N6 in the lower four. The shape of the mean buoyancy does not change significantly for run N6 up to the cloud boundary as seen in panels (e), (f) and (g). Above the cloud base at $z\approx 0.5$  it appears practically with reversed maxima and minima as seen in panel (h). The same is observed for run D6 in Fig. \ref{Bild12} when panels (b) and (d) are compared. The reversion of maxima with minima is seen to take place in panel (c). This effect is clearly a result of evaporation and condensation.

\begin{figure}
\includegraphics{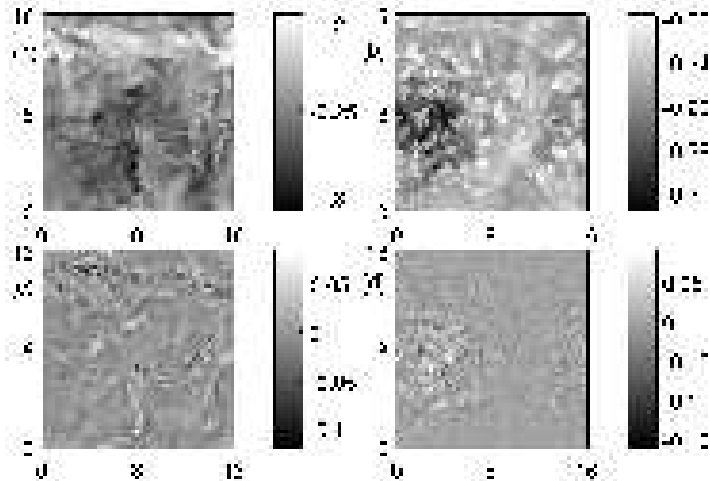}
\caption{Snapshot of buoyancy and vertical velocity at height $z=0.3087$. (a) $B$ from N6, (b) $B$ from D6, (c) $u_z$ from N6, (d) $u_z$ from D6.}
\label{Bild13}
\end{figure}

In Fig. \ref{Bild13} we show snapshots of the buoyancy $B$ and the corresponding vertical velocity $u_z$ in the lower stably stratified layer close to the bottom plane. The figure displays an anticorrelation of buoyancy and vertical velocity in the dry layer which was already found in Fig. \ref{Bild6} as a negative mean convective buoyancy flux $T_{conv}(B)$. In this figure it can  be seen how it arises. Downwelling plumes have a higher buoyancy than the ambient air. For both sets of boundary conditions this seems to be similar and we give a brief explanation. If a parcel of unsaturated air starts to ascend from the inner part of the lower dry layer fluid has to move downward in its vicinity due to incompressibility. Such ascending air parcels can be less buoyant than the surrounding ambient air. When they come to a height where the saturation threshold is reached latent heat can be released due to condensation of liquid water. This causes the reversal of buoyancy which is seen in Fig. \ref{Bild12}. Now upward motion of the 
parcel gets amplified and eventually the upper boundary plane (see Figs. \ref{Bild8}(a) and (b)) can be reached. Incompressibility forces the parcel to sink back into the cloudy bulk, eventually even to penetrate the interface between cloudy and cloudless layer and to fall from above into the stably stratified lower part. As we mentioned already earlier this vertical exchange becomes increasingly difficult with increasing Rayleigh number and thus strongly prevents convective transport of heat and moisture across the cloud boundary.
%--------------------------------------------------------------------------------
%--------------------------------------------------------------------------

\section{Conclusions and outlook}

We studied moist turbulent Rayleigh-B\'{e}nard convection for two different boundary conditions of the dry and moist buoyancy fields by means of the three-dimensional direct numerical simulations. These are either a fixed amplitude boundary condition at the top and bottom plane or a fixed flux boundary condition. For smaller Rayleigh numbers differences in the behavior of the convection are observable as our analysis of vertical mean profiles of buoyancies, velocity fluctuations and buoyancy transport currents shows. Differences in the velocity and buoyancy fields occur due to the fact that in the case of Neumann BC the scalar fields can vary at the plates. With increasing Rayleigh number these differences decrease such that at $Ra_M^*=10^7$ statistical properties are nearly the same. 

A comment on the velocity boundary conditions is in order here: Many of the cited references use no-slip BC for the velocity field instead of free-slip BC. We also performed some simulations with these BC in order to see if there are essential differences. For the Rayleigh numbers studied, we found no significant differences. The qualitative picture remained unchanged. It is clear that flow patterns and mean properties will be affected, in particular close to the boundaries where the horizontal components have to be zero then as well. A further aspect was that we wanted to compare our studies with the previous work on this subject such as those by Kuo\cite{Kuo} and Bretherton.\cite{Bretherton1,Bretherton2} In terms of a more realistic configuration with respect to the atmospheric dynamics mixed BC (no-slip at the bottom and free-slip at the top) would be a possible case which is an interesting aspect of future work.

For the present parameter settings, the convection layer can be divided into two subregions. In the upper part of the convection layer a closed cloud layer is found in which the saturated air parcels are in turbulent motion. The lower part consists of a layer of unsaturated air which is mostly stably stratified and inhibits significant convective motion as indicated by a significant reduction of the turbulent velocity fluctuations and the convective buoyancy transport. In this layer diffusive transport of moisture and heat dominates. Since the mean diffusive equilibria are the same for both sets of boundary conditions  similar mean profiles result particularly for the higher Rayleigh numbers. Our DNS indicate that with increasing Rayleigh number the vertical transport across the full layer is more and more diminished. One reason for this behavior is that the diffusive transport in general gets less and less efficient to sustain fluid motion in the dry part. In a nutshell, the convection in both subsystems decouples increasingly of each other. In the upper subsystem, we observe classical Rayleigh-B\'enard convection of fully saturated air, in the lower part weak fluid motion in a stably stratified environment. The interfacial region between both subsystems in which phase changes are present becomes narrower which is indicated by a decrease of the variance of the cloud base, but also by the shaper bump in the mean profiles of the Nusselt number $Nu_B(z)$ in Fig. \ref{Bild6}. One resulting question is which additional physical processes can re-amplify the vertical transport across the whole layer and break up this separation between the sublayers? A suggestion is to add radiative transfer or precipitation processes.

Although the qualitative behavior of moist turbulent convection in the conditionally unstable regime differs significantly from the classical dry convection case, one main outcome of the study in the present MRBC model is very similar to what has been found in dry Rayleigh-B\'{e}nard convection. With increasing Rayleigh number the global transport properties for runs with both sets of boundary conditions converge to each other. This finding seems to be also robust when switching from the two-dimensional case as in Ref. \onlinecite{Johnston} to three dimensions as here. Although the structures of the turbulence fields in the three-dimensional case are much more complex some basic mechanisms of the upward transport of heat (and moisture) are similar.

The properties which are relevant for atmospheric processes such as  the cloud fraction or the vertical transport of heat and moisture were not affected by the particular choice of the BC in our MRBC model for the highest accessible Rayleigh number. As mentioned already above, interesting would be therefore to repeat such study with further physical processes included, such as the radiative transfer and precipitation in order to test if this insensitivity persists. This has been started in part\cite{Pauluis3} and will be continued in the future.

\begin{acknowledgments}
This work is supported by the Deutsche Forschungsgemeinschaft (DFG) under Grant No. SCHU1410/8-1 and the DFG Heisenberg Program under Grant No. SCHU1410/5-1. We thank Olivier Pauluis and Siegfried Raasch for discussions. Support with computer time at the J\"ulich Supercomputing Centre under Grant HIL02 is also acknowledged.
\end{acknowledgments}

\end{document}